\documentclass[12pt]{iopart}
\usepackage{epsf}
\usepackage{graphicx}
\begin{document}

\title[Spin Glasses: Model systems...]{Spin Glasses: Model systems for non-equilibrium dynamics}

\author{Per Nordblad}

\address{Department of Materials Science, Uppsala University, Box 534, SE-751 21 Uppsala, Sweden}
\ead{per.nordblad@angstrom.uu.se}

\begin{abstract}
Spin glasses are frustrated magnetic systems due to a random
distribution of ferro- and antiferromagnetic interactions. An
experimental three dimensional (3d) spin glass exhibits a second
order phase transition to a low temperature spin glass phase
regardless of the spin dimensionality. In addition, the low
temperature phase of Ising and Heisenberg spin glasses exhibits
similar non-equilibrium dynamics and an infinitely slow approach
towards a thermodynamic equilibrium state. There are however
significant differences in the detailed character of the dynamics
as to memory and rejuvenation phenomena and the influence of
critical dynamics on the behaviour. In this article, some aspects
of the non-equilibrium dynamics of an Ising and a Heisenberg spin
glass are briefly reviewed and some comparisons are made to other
glassy systems that exhibit magnetic non-equilibrium dynamics.

\end{abstract}


\submitto{\JPCM}

\maketitle

\section{Introduction}
\begin{figure}
\begin{center}
\includegraphics[width=4.0in]{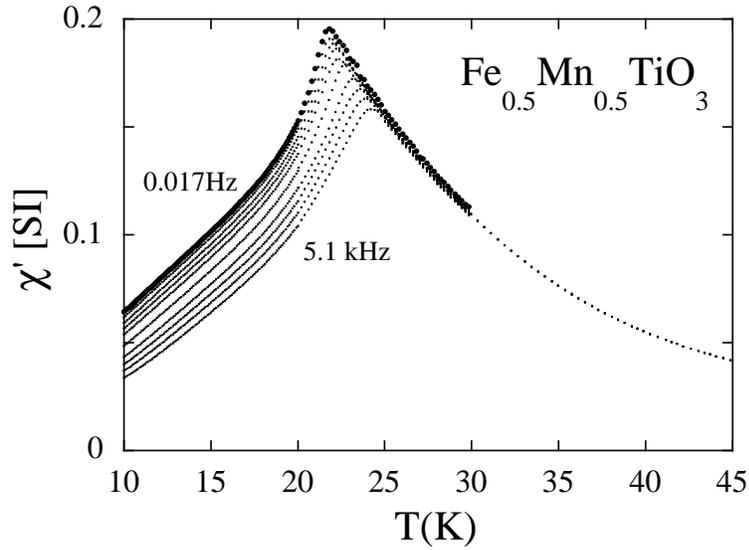}
\end{center}
\caption{\label{Ising}The ac-susceptibility of
$Fe_{0.5}Mn_{0.5}TiO_3$ at logarithmically evenly spaced
frequencies from 0.017 Hz to 1.7 kHz (top to bottom).}
\end{figure}

\begin{figure}
\begin{center}
\includegraphics[height=3.0in]{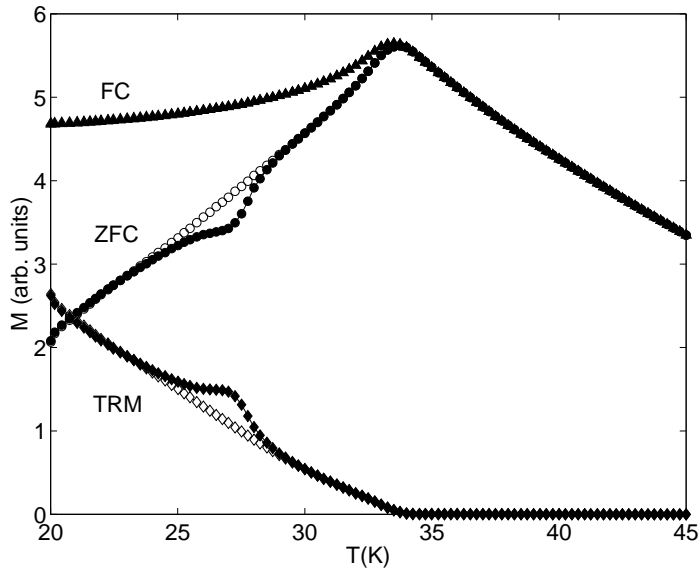}
\end{center}
\caption{\label{Heisenberg}The ZFC, FC and TRM magnetisation of
Ag(11at\%Mn) measured on continuous heating at the same rate after
having cooled the system continuously (open circles)and after
cooling it with stops for some hours at 27 K (filled circles).
Reproduced from reference \cite{Math}.}
\end{figure}

Dilute magnetic alloys exhibit a second order phase transition at
concentrations of magnetic constituents well below the percolation
limit. This transition - the spin glass transition - has since its
discovery \cite{Myd} caused a lot of research activities and
provided us with new magnetic phenomena. The conspicuous sharp
cusp and frequency dependence of the ac susceptibility is
exemplified in figure \ref{Ising} by measurements on the 3d Ising
system $Fe_{0.5}Mn_{0.5}TiO_3$. In low field dc magnetisation
experiments, the spin glass transition is revealed by a maximum in
the zero field cooled (ZFC) magnetisation, an irreversibility
between the ZFC and the field cooled (FC) magnetisation and a
continuous decay of the thermo remanent (TRM) magnetistion to zero
at the temperature where the irreversibility between the ZFC and
FC magnetisation appears (see figure \ref{Heisenberg}). Static and
dynamic scaling analyzes of the critical behaviour support the
occurrence of a second order phase transition for both 3d Ising
and Heisenberg spin glasses. As to the dynamics, it is of certain
interest to note that in contrast to ordinary ferro- or
antiferromagnetic phase transitions, the character of the spin
glass transition allows experimental determination of the critical
slowing down exponent, $z\nu$, on the time scales of low frequency
ac-susceptibility experiments. In addition, the low temperature
spin glass phase owns intriguing non-equilibrium dynamics, which
is the main subject of the current article.

\section{Critical dynamics}

Experimentally it has been amply demonstrated that both Ising and
Heisenberg systems exhibit dynamic critical behavior on
approaching the spin glass temperature $T_g$. Analyzes of data
from low frequency ac-susceptibility experiments indicate a
divergence of the relaxation time as the glass temperature is
approached, with a critical exponent, z$\nu$, that is of order
8-10. However, when comparing results from different experimental
systems, it is yet not possible to distinguish significantly
different values of the exponents on systems of different spin
dimensionality. To illustrate the behaviour we quote analyzes of
the slowing down of the dynamics close to $T_g$ of the model Ising
system $Fe_{0.5}Mn_{0.5}TiO_3$.\cite{Gun,Mat} The result of these
studies is that the dynamic exponent z$\nu$ should be about 10 for
the 3d Ising system. Quite similar values are also obtained from
analyzes of measurements on dilute magnetic alloys (Heisenberg
systems) as well as on strongly interacting magnetic nano particle
systems \cite{Han} (super spin glasses \cite{Kle}). It is striking
to learn from these experiments that the disordered and frustrated
spin glass allows studies of critical behavior at time scales and
reduced temperatures that for a pure ferro- or antiferromagnetic
system would reach far away from the critical region. One could
then expect that the critical behaviour of spin glasses should
yield very reliable and unambiguous results as to the values of
the critical exponents and possible differences between systems of
different universality class. This is however not the case, due to
strong limitations in how close to the transition it is possible
to perform meaningful experiments, the slowing down of the
dynamics and entrance into nonergodic regions limit the studies to
only order of 0.01 in reduced temperature. The conclusion from
dynamic scaling analyzes is that there is a spin glass transition
in zero applied magnetic field, but is this also true in an
applied magnetic field?

\subsection{In-field criticality?}

\begin{figure}
\begin{center}
\includegraphics[width=4.0in]{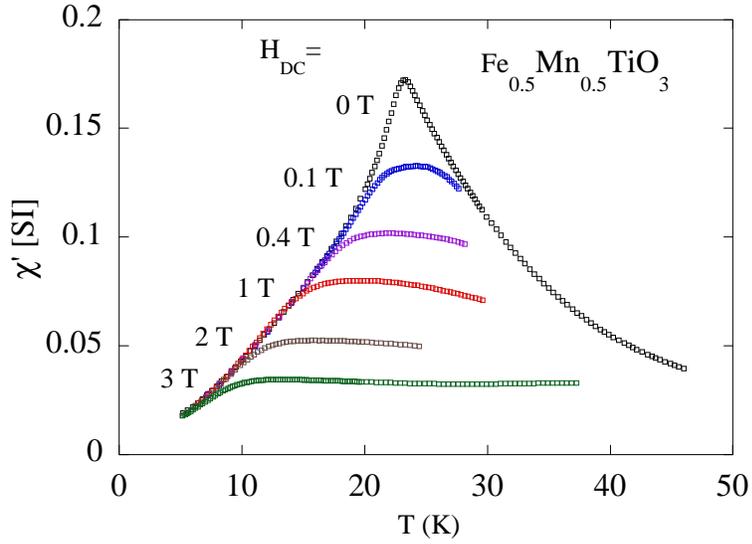}
\end{center}
\caption{\label{ac-H}The ac-susceptibility of
$Fe_{0.5}Mn_{0.5}TiO_3$ measured in an weak 125 Hz ac-field in
superposed dc magnetic fields as indicated in the figure.}
\end{figure}

To understand spin glasses one needs a useful and predictive
modelling of the systems. There are two main approaches to
describe spin glasses - starting from phase space (mean field
models) and real space (droplet scaling models). The mean field
model predicts a finite spin glass temperature and a persistence
of the phase also in applied magnetic fields, the spin glass and
the paramagnetic phase separated by the AT-line \cite{Alm}. The
droplet scaling theory of 3d Ising spin glasses on the other hand
predicts that in the thermodynamic limit, any finite magnetic
field destroys the spin glass phase, i.e. the in-field
thermodynamic equilibrium phase is paramagnetic. Within this
picture, this does not imply that critical dynamics cannot be seen
in weak enough magnetic fields. A crucial point in this model is
the correspondence between time and length scales, i.e an
experimental probe that measures at a certain frequency or time
scale also probes the system on a length scale set by the
observation time (and the temperature). A finite field sets an
upper limit to the correlation length in the spin glass: On
shorter length scales the system appears unaffected by the field
and on larger length scales the system is at equilibrium
(paramagnetic). The ac susceptibility measured at low fields is
dynamically limited by the frequency, $\omega$, of the varying ac
field and the susceptibility (at observation time, t=1/$\omega$)
increases with decreasing frequency. At temperatures above the
freezing temperature (cusp in the real part of the
susceptibility), the susceptibility becomes frequency independent
on lowering the frequency. When a dc field is superposed on the
system, the ac-susceptibility is unaffected if the field is weak
enough, however after a certain field magnitude, the
susceptibility cusp becomes suppressed and the freezing
temperature is pushed to lower temperatures. However, at lower
temperatures, the ac-susceptibility remains unaffected, i.e. one
cannot distinguish the in field from the zero field data (see
figure \ref{ac-H}). An analysis of the in-field slowing down of
the dynamics of the 3d Ising spin glass $Fe_{0.5}Mn_{0.5}TiO_3$
suggests that critical slowing down does not describe the behavior
and that the spin glass transition is destroyed by any magnetic
field.\cite{Mat} It should in this context be mentioned that a
careful study using torque magnetometery of the spin glass
transition on systems with varying degree of anisotropy is
supportive of the existence of an in-field phase transition for
Heisenberg spin glasses.\cite{Dor} Such a difference between the
in-field behaviour of Ising and Heisenberg systems is of course
remarkable, and is not expected within a droplet scaling picture
of the spin glass.

\section{Non-equilibrium dynamics}

At temperatures below the spin glass temperature, the zero field
spin glass never reaches equilibrium, the equilibration time is
infinite. Experiments on this phase are thus looking at a
non-equilibrium system and the results are age dependent. The
non-equilibrium character can be experimentally observed from an
age dependence of the magnetic response.\cite{Lun}  In certain
temperature ranges the aging imposes a magnetic response of
instructive and almost simplistic character. There is an
inflection point in a plot of the magnetisation ($M$) relaxation
vs log $t$ at an observation time equal to the wait time, $t_w$,
before the application (or removal) of a weak magnetic field, and
the relaxation rate ($S(t)=(1/H)dM(t)/dlog t$) exhibits a maximum
at the same observation time. This behaviour reveals much of the
physics behind the aging phenomenon in spin glasses and together
with results from experiments according to somewhat altered
thermal protocols it has been possible to relate effective aging
times to the sizes of equilibrium spin glass domains. This
length-time scale (droplet model) approach to the spin glass
problem will be adopted below to interpret the non-equilibrium
dynamics as measured in low frequency ac-susceptibility and
dc-magnetic relaxation experiments.

The time scale for spin reversal of an atomic spin is about
$10^{-13}$ in spin glasses. On approaching the glass temperature,
correlation between spins occurs on longer and longer length
scales and the response time of the system diverges at $T_g$. At
lower temperatures the response time again decreases according to
critical slowing down, however, the random spin structure that is
obtained after a quench is not compatible with the competing
interaction pattern, and there is a slow restructuring of the
system on larger and larger length scales. In terms of the droplet
scaling model this is governed by thermal activation over an
energy barrier that increases with decreasing temperature and the
growth occurs logarithmically in time.\cite{Fis} However, the
equilibrium configuration at one temperature is different from the
equilibrium structure at any other temperature (temperature
chaos).

\subsection{Isothermal aging}

\begin{figure}
\begin{center}
\includegraphics[width=3.0in]{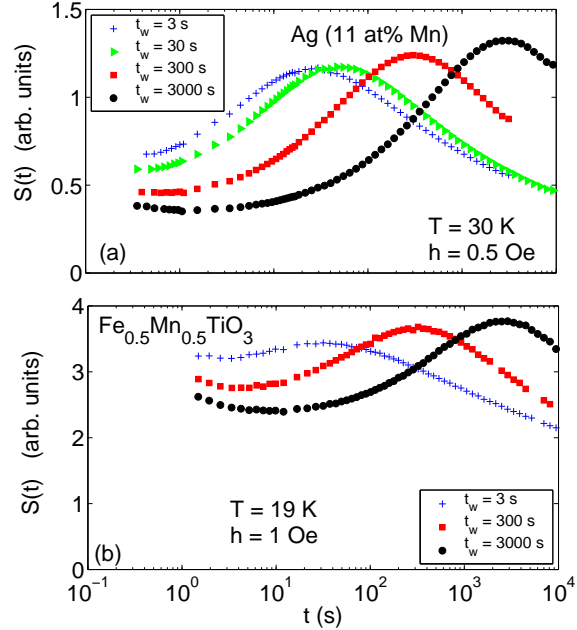}
\end{center}
\caption{\label{Rate}Relaxation rates for the ZFC relaxation of
Ag(11\%Mn) and $Fe_{0.5}Mn_{0.5}TiO_3$ after different wait times
at the measurement temperature. The experiments are in both cases
performed at $T/T_g$=0.9. \cite{Jonp2}}
\end{figure}
 Quenching the system from a temperature above $T_g$ to a measurement temperature
 may seem to be a simple procedure, but experimentally, quenching in a
 magnetometer implies cooling rates of only some kelvin per minute. This is
 of course a slow cooling compared to a spin flip time of only  $10^{-13}$ s.
 We are however fortunate in the sense that also experiments in a magnetometer
 system are confined to long time scales compared to the spin flip time.
 Figure \ref{Rate} shows results from an isothermal aging experiment on an
 Ising and a Heisenberg system, plotted in the figure is the magnetisation
 relaxation and the relaxation rate vs. time after the field application in
 a semi-logarithmic diagram. Both systems show a maximum relaxation rate
 at a time of the order wait time before applying the magnetic field. A striking
 difference between the behaviour of the two systems is that the aging has a
 much larger influence on the relaxation of the Heisenberg system than of
 the Ising system. In the measured time window, the relaxation rate rises
 a factor of about 5 above an estimated equilibrium relaxation rate in the Ag(Mn)
 spin glass but only a factor less than 1 for the Ising system. Additionally,
 for both systems it does not matter at what temperature we choose to
 make our measurement,
 an aging behaviour is always observed. In fact, provided the
 temperature step
 is large enough, one can make two sequential stops at a higher and a lower
 temperature, and the observed aging at the lower temperature is the same as
 if the system was immediately cooled from above $T_g$ without the stop. The
 aging is non-accumulative in both systems, however a larger separation
 between the two temperatures is required for the Ising than the Heisenberg system.

\subsection{Memory}
The non-accumulative nature of the aging and a preserved memory of
the cooling process is best illustrated by a zero field cooled
magnetisation vs temperature experiment under two different
cooling protocols. The behaviour was introduced already in figure
2, where the low field ZFC, FC and TRM magnetisation of the Ag(Mn)
sample was plotted vs. temperature. The protocol including a stop
at a certain temperature shows in the ZFC case a rather narrow dip
that has a logarithmically increasing depth with increasing stop
time. A corresponding, but rather more shallow and broader dip is
observed in experiments on the Ising sample. A memory of the spin
structure attained at the stop temperature is imprinted and
preserved during the succeeding cooling-heating process. The
understanding of the non-equilibrium dynamics may be further
elucidated by low frequency ac-susceptibility using the same
thermal protocol as in the here described dc-magnetisation
studies. Using such studies, the temperature memory behaviour
\cite{Ham} was first explicitly reported by Jonason et al
\cite{Jonk} in measurement of the out of phase component of the
ac-susceptibility vs temperature on a $CdCr_{1.7}In_{0.3}S_4$ spin
glass.

Using the functional form of the decay of the out-of-phase
component of ac-susceptibility under isothermal aging at different
temperatures and frequencies, a quantitative measure of the
increasing spin glass domain size in a Ag(Mn) spin glass in terms
of the droplet scaling model has recently been derived by
J\"onsson et al \cite{Jonp1}. One ingredient of the description of
the aging behaviour that these and other experimental results
imply is the existence of an overlap on short length scales
between a spin glass domain structure developed at one temperature
and the domain structure at a different temperature $T_m + \Delta
T$. The concept accounting for this is the overlap length
\cite{Bra}, the validation and influence of which can be further
studied by experiments after applying perturbations (temperature
steps or cyclings) to the system after an aging period.

\subsection{The overlap length and ghost domains}
\begin{figure}
\begin{center}
\includegraphics[width=3.0in]{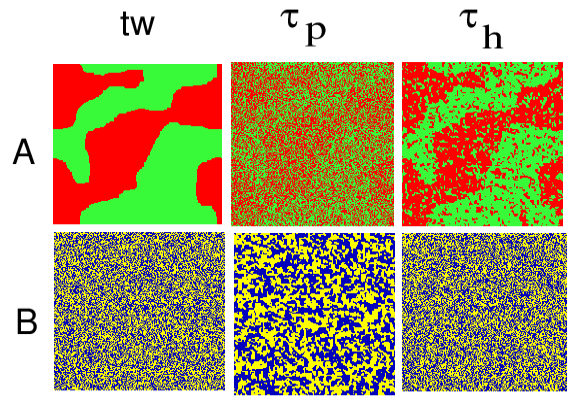}
\end{center}
\caption{\label{ghost}The evolution of domains in a 2d Mattis
model \cite{Yos} during a bond perturbation experiment. The system
is allowed to evolve under two different interaction patterns
having completely different ground states (A and B). The spin
structure of the system is projected onto the ground states A and
B at three different stages of a perturbation simulation. Left
figures, after the first wait time under interaction pattern A:
Grown plus and minus domains are seen in the projection on A,
whereas the projection on B yields a random pattern. Middle
figures, after the perturbation under conditions B, a domain
pattern has evolved in the projection onto B; and in the
projection onto A one can still resolve the sign and size of the
original domains (ghosts), however in a lot of random noise. Right
figures, after evolving under the interaction pattern A again, the
noise on the projection onto A is largely washed away and the
original domain structure recovered. Reproduced from
\cite{Jonp2}.}
\end{figure}

There is a quite extensive literature on the influence of
perturbations (temperature, field or bond) on the spin glass
dynamics both from the point of view of experiments and
simulations. One recent theoretical development of the droplet
model introducing the ghost domain concept \cite{Yos} provides a
visual illustration of how an original domain structure is
recovered after a perturbation that intuitively should destroy any
memory of the preceding aging process. The procedure and process
are illustrated in figure \ref{ghost} for bond perturbations on a
2d Mattis model. In terms of a temperature cycling experiment, the
important features are: The system is first aged at a constant
temperature and spin glass domains of "up" and "down" sign grow.
After a wait time at this temperature, the temperature of the
system is shifted and the system is allowed to evolve under these
new conditions. At this temperature, domains starts to grow
starting from the overlap length and having a completely different
structure than the original domains on larger length scales. The
temperature is then shifted back to the original one. However, at
this temperature, a (ghost) pattern of the original domain
structure remains and the perturbation is seen as patches of the
wrong domain kind within the original domains. Keeping the system
at this temperature, the patches are gradually washed away and
simultaneously the original domains continue to grow. The
experiments that reflect and promote that these kind of internal
processes occur are magnetic relaxation and relaxation of the ac
susceptibility recorded after the perturbation protocols.
\cite{Jonp2}

\section{Conclusions}
The Non-equilibrium dynamics of spin glasses is generic to both
Ising and Heisenberg systems, however with some significant
differences in the detailed behaviour. In a recent article by F.
Bert et al \cite{Bert} a systematic study of the effects of
perturbations on the aging behaviour of spin glasses with
different anisotropy ranging from purely Ising to closely
isotropic systems was reported. Also other magnetic systems show
spin glass like non-equilibrium dynamics; characteristic of Ising
systems is the critical dynamics \cite{Han} and memory behaviour
\cite{Jonp3} of a near mono dispersed interacting nano particle
system, whereas certain features of the aging and memory behaviour
of a high temperature superconductor showing the paramagnetic
Meissner effect \cite{Gar} are more Heisenberg like.

\ack{Financial support from the Swedish Research Council (VR) and
the European Union programme RTN through the network DYGLAGEMEM is
acknowledged. Recent collaboration with Petra J\"onsson, Roland
Mathieu and Hajime Yoshino has been especially important for the
current article.}

\end{document}